# Atomic Scale Investigation of Dye Sensitized Solar Cells: Interface Structure and Dynamics

*Wei Ma, Fan Zhang and Sheng Meng\**

[1]Beijing National Laboratory for Condensed Matter Physics, and Institute of Physics, Chinese Academy of Sciences, Beijing, 100190, P. R. China

[2]Collaborative Innovation Center of Quantum Matter, Beijing, 100190, P. R. China

**ABSTRACT**

Dye-sensitized solar cells (DSC), built upon organic/inorganic hybrid materials, are possibly the most promising third generation solar cells, which attract wide attentions due to their low-cost, flexibility and environment-friendly properties. We review recent progresses in DSC research, focusing on atomic-scale investigations of the interface electronic structures and dynamical processes, including dye adsorption structure onto $TiO_2$, ultrafast electron injection, hot-electron injection, multiple-exciton generation and electron-hole recombination. We briefly summarize advanced experimental techniques and theoretical approaches investigating these interface electronic properties and dynamics, and then introduce progressive achievements in the photovoltaic device optimization based on insights obtained from atomic scale investigations. Finally, some challenges and opportunities for further improvement of dye solar cells are presented.

\*E-mail: smeng@iphy.ac.cn.

## 1. INTRODUCTION

Since the first practical photovoltaic cell, which was based on a diffusive silicon p–n junction and reached an efficiency of 6%, was developed in 1954 at Bell Laboratories by Daryl Chapin et al.,[1] the solar cells technology have come to a new age. Dye-sensitized solar cells (DSC), one of the most promising third generation solar cells, based on highly porous nanocrystalline titanium dioxide films and organic dyes, have drawn considerable technological interest for their potential to decrease manufacturing costs and for their demonstrated high power energy conversion efficiency, since the seminal work by Grätzel et al. in 1991.[2] The highest solar-to-electricity power conversion efficiency (PCE) for molecular DSC is 13% under AM1.5G full sun irradiation, obtained by sensitization of modified zinc-porphyrin-based donor-π-acceptor (D-π-A) dye in 2014.[3] Wang et al. have achieved comparable efficiency of 12.8% at *half* irradiance of the AM1.5G sunlight using metal-free all-organic dyes, which possesses large molar absorption coefficient, benign environment impact and low cost, compared with dyes containing heavy metals.[4]

The current DSC consists of different layers of components stacked in serial, including transparent conducting glass substrate, transparent conducting layer, $TiO_2$ nanoparticles, dyes, electrolyte, and counter electrode covered with sealing gasket. The typical configuration is shown in Fig. 1a. As shown in Fig. 1b, the operation of a dye-sensitized solar cell starts with the photo-excitation of the sensitizer, where an electron is excited from the ground state to higher-energy excited states of the sensitizer by photon absorption. Then the excited electron injects rapidly from the photo-excited sensitizer molecule to the conduction band of the semiconductor, leading to the formation of mobile electrons (and a dye cation). On one hand, titanium dioxide, acting as electron-transport material, transports the injected electrons to the back conductive contact. On the other hand, electrolyte, serving as hole-transport material, reduces oxidized dyes and transports holes to the counter electrode.

In 1961, Shockley and Queisser calculated the maximum theoretical solar conversion efficiency of a solar cell using a single p-n junction. They found the maximum efficience is 33.7% under standard AM1.5G solar irradiation, reached with a band gap of 1.37 eV, known as Shockley-Queisser limit or detailed balance limit.[5] According to Shockley-Queisser limit, energies are lost mainly through the following four ways in DSC:

  i. Blackbody radiation, which is a type of electromagnetic radiation within or surrounding a solar cell in thermodynamic equilibrium with its environment, representing 7% of the available incoming solar energy.
  ii. Spectrum losses. Only photons with energy higher than the HOMO-LUMO band gap can be absorbed by the sensitizers in DSC, which means only ultraviolet and visible light will contribute to power production, whereas infrared, microwaves and video waves will not.
  iii. Thermal relaxation, which contains two processes: (a) electrons in the excited states of the chromophores easily jump back to ground state if not rapidly inject to $TiO_2$ conduction band, (b) injected electrons in semiconductor conduction band tend to thermally relax to the conduction band edge rapidly.
  iv. Radiative recombination. Electrons in $TiO_2$ conduction band will recombine with holes

in electrolyte and dyes if not efficiently transfer to the conducting contact, causing power loss by emitting photons.

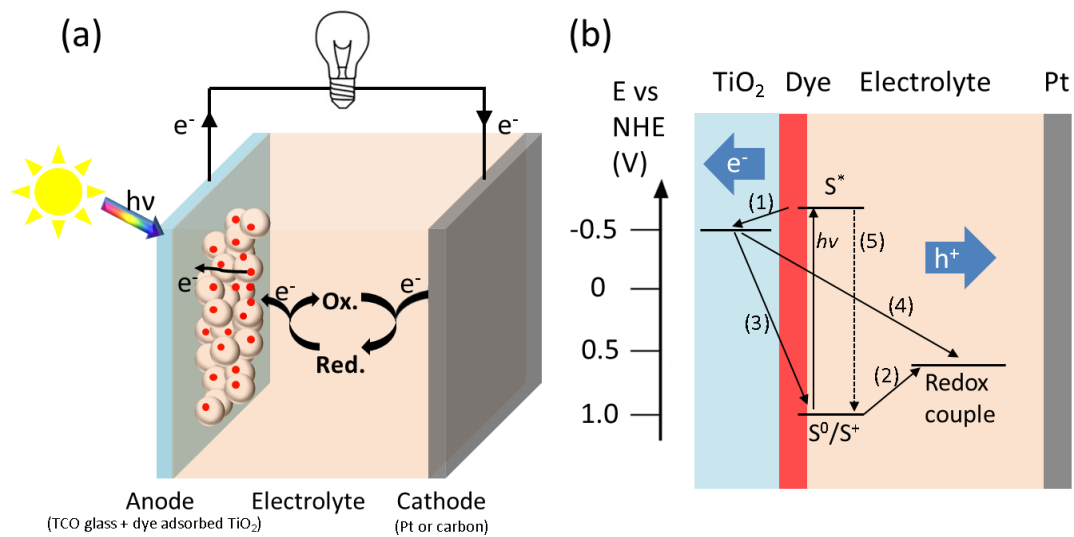

Figure 1. (a) General scheme of dye-sensitized solar cells. (b) Typical electronic dynamic processes in the dye-sensitized solar cells.

Among all the four main pathways causing loss of incoming energy, blackbody radiation is inevitable. Therefore, reducing spectrum losses, thermal relaxation and radiative recombination are major ways to optimize dye-sensitized solar cell. In this review, we broke down energy conversion in DSC into individual processes taking place at solar cell interfaces, including dye/$TiO_2$ interface optimization, dye absorption properties, electron injection, thermal relaxation and electron-hole recombination. In section 2, we mainly introduce experimental and theoretical methods to identify the precise interface structure of dye/$TiO_2$ system, discussing the adsorption structures of organic chromophores on the semiconductor substrate, and presenting advanced methods to control the dye/$TiO_2$ binding configurations. In section 3, the energy gap optimization to improve photon absorption is presented. Frequently used methods to reduce the solar spectrum losses, including co-sensitization of two sensitizers with complementary absorption spectra and using near infrared dyes, are discussed. In section 4, we briefly introduce advanced experimental techniques and theoretical approaches to investigate the interface electron transfer dynamics, and factors influencing electron injection rates at dye/$TiO_2$ interface. Hot-electron injection and multiple-exciton generation are presented in section 5 as two efficient novel ways to reduce thermal relaxation losses. In section 6, a short summary and future prospective are given.

## 2. ADSORPTION STRUCTURE
### 2.1 Identification methods

In a DSC, dye adsorption is the first basic and important step for power production. Only when dye molecules bind effectively to the semiconductor substrate can the following processes such as electron injection and charge transportation proceed with high efficiency. Knowing the precise interface structures upon dye adsorption onto $TiO_2$ is of crucial importance for further device optimization. Infra-red (IR) spectroscopy and Raman spectroscopy are mostly used spectral techniques to investigate the adsorbed layer and anchoring modes in experiment.

IR spectroscopy is applied to determine the nature of the adsorption groups and the mode of their interaction with the substrate, the changes caused in the adsorbed molecule by the field of the $TiO_2$ substrate, and the nature of new chemical compounds and/or bonds formed upon adsorption.[6] The theory of IR techniques shows that a molecule, as a whole, carries out so-called vibrations, in which the amplitude of motion differs from different atoms, while all atoms vibrate at the same frequency. When the amplitude of one of the vibration modes is considerably greater than that of the others, it become the characteristic vibration of the particular bond or groups of atoms (-$CH_3$, > $CH_2$, > CO [6]). Fourier transform infrared (FTIR) spectroscopy is a measurement technique to record infrared spectra, which is widely used in experiments to recognize the adsorption structure of dye on metal oxides with the help of the Deacon-Philips rule.[7] Taking carboxylate dyes as an example, an important parameter $\Delta v$, which is defined as the frequency splitting of the asymmetric and symmetric vibrations of surface bound carboxylate, is measured. Two cases are compared: $\Delta v$ for dyes in solid state, $\Delta v(solid)$, and $\Delta v$ for the adsorbed dyes, $\Delta v(ads)$. If $\Delta v(ads) > \Delta v(solid)$, the dye molecule takes a monodentate binding mode; if $\Delta v(ads) < \Delta v(solid)$, the bidentate bridging mode is more preferred; if $\Delta v(ads) \ll \Delta v(solid)$, the chelating mode is most likely to present.[7]

However, IR spectroscopy is inactive in homonuclear diatomic molecules and complex molecules whose vibrational modes are weak in, or apparently absent from the IR spectrum.[8] Fortunately, Raman spectroscopy offers distinct advantages in detecting and analyzing molecules with inactive IR spectra. Moreover, Raman spectra can be employed to study materials in aqueous solution, a medium that transmits IR poorly.[8] Therefore sample preparation for Raman is generally simpler than for the IR measurement. Raman technique relies on inelastic scattering, or Raman scattering, of monochromatic light, usually from a laser in the visible, near infrared, or near ultraviolet range. The laser light interacts with molecular vibrations, phonons or other excitations in the system, resulting in the energy of the laser photons being shifted up or down. The shift in energy gives information about the vibrational modes in the system under study.

Other methods, such as nuclear magnetic resonance (NMR),[9] Auger electron spectroscopy (AES),[10-11] high-resolution electron energy loss spectroscopy (HREELS),[12-13] ultra-violent photoelectron spectroscopy (UPS),[14] X-ray photoelectron spectroscopy (XPS),[15] temperature programmed desorption technique (TPD)[16] and thermal desorption spectroscopy (TDS),[17] are also useful for identification of interface structures and can provide more information of the intricate dye/$TiO_2$ interface adsorption phenomenon. Theoretical methods often confirm the interface binding configurations by comparing binding energies of different adsorption structures.[18]

**2.2 Anchor groups**

For the majority of metal complex dyes, a carboxyl group is employed as an effective anchor through which dyes bind onto $TiO_2$ surfaces. Experimental[19-20] and theoretical analyses[21] revealed that N719 and derivatives bind onto anatase (101) via one to three carboxylic/carboxylate groups forming bidentate or monodentate for each group.

For all-organic dyes, cyanoacrylic group[22-23] and phosphoric acid group[24-25] are commonly used as binding unit. Phosphoric acid groups are known to adsorb strongly to most metal oxides and adsorb on the $TiO_2$ surface via a bidentate binding of phosphonate to Ti(IV) ions by in situ internal reflection infrared spectroscopy.[25] Cyanoacrylic acid groups are mostly used as anchoring moiety in all-organic donor-π-bridge acceptor dyes, combining the electron withdrawing properties of the cyano-unit with the binding motif of the carboxylic group.[22-23] However, there exists controversy concerning the adsorption configurations of cyanoacrylic dyes. Intuitively, it is widely assumed that all-organic cyanoacrylic dyes also bind the $TiO_2$ surface through their carboxylic group, similar to N719. In 2007, Johansson et al. indicated that L2 dye adsorbed onto $TiO_2$ surface with a dominating orientation that the diphenylaniline donor moiety pointing out from the surface [26]. In 2012, Jiao et al. proposed a tridentate anchoring site of all-organic cyanoacrylic dye featuring Ti-N bonding in DSC based on first-principles molecular dynamics and real-time time dependent density functional theory. [18] The cyano group, not only acts as an electron-drawing acceptor, but also directly binds onto $TiO_2$ and contributes to interface stability. As shown in figure 2a, numerous adsorption configurations of model cyanoacrylic M0 dye on prototypical $TiO_2$ anatase (101) surface are considered. Extensive energetic, vibrational recognition and electronic data revealed that Ic is the most stable configuration with optimal energy alignment to minimize kinetic redundance and presenting ultrafast photoelectron injection dynamics with high yield. In a recent work, a novel acyloin anchor group is found to strongly bind to $TiO_2$ semiconductors and enable efficient electron injection into substrate. [27] However, the detailed interface configuration is still unknown.

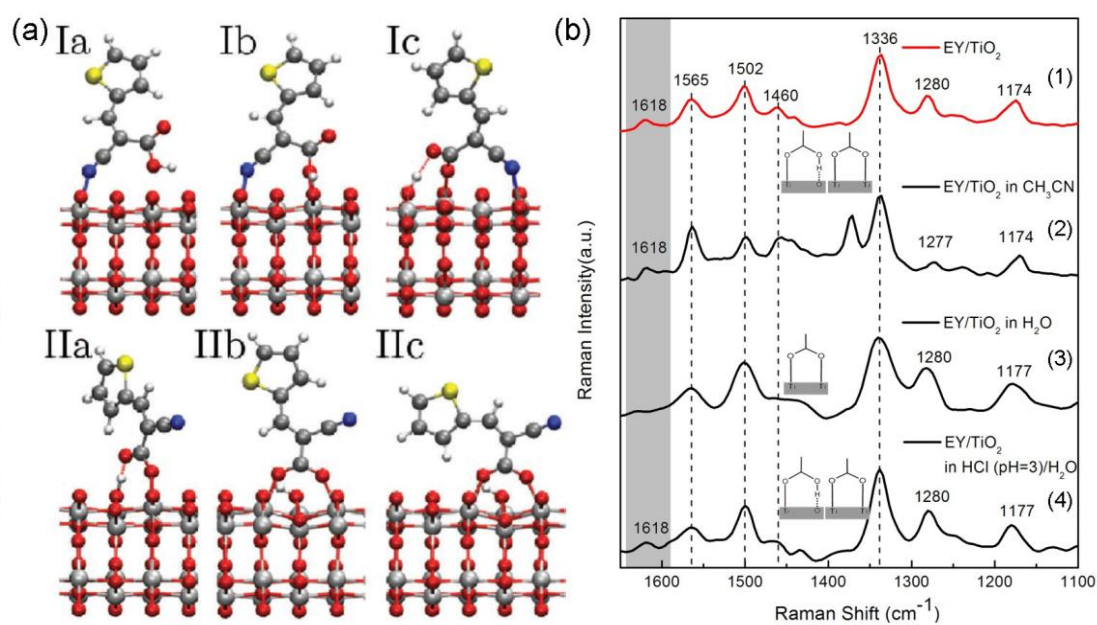

Figure 2. (a) Different adsorption structures of the M0 model dyes binding on $TiO_2$ anatase (101) surface. First row: adsorption with Ti-N bonding. Second row: adsorption without Ti-N bonding. [18] (b) Raman spectrum of Eosin Y adsorbed $TiO_2$ under different conditions: (1) in air, (2) in acetonitrile, (3) in pure water, and (4) in HCl aqueous solution at pH 3. [28]

**2.3 Manipulation of interface adsorption structures**

With solid and comprehensive characterization of the dye/$TiO_2$ interface geometry, a complete understanding of the working mechanism of DSC and details about interface electronic structure and dynamics can be achieved. De Angelis et al. investigated the adsorption configuration of most popular Ru-complex N-719 dye on $TiO_2$ and showed that the dipole moment orientation of the sensitizers resulted from different binding configurations, which can lead to as large as a 0.61 eV shift in $TiO_2$ conduction band edge (CBE), introducing a larger open circuit voltage ($V$oc). [21] Jiao et al. studied the adsorption configuration of all-organic cyanoacrylic dyes and found that interface adsorption structure with Ti−N binding is beneficial to electron injection, which improves short circuit current ($J_{SC}$).[18]

Interface binding configurations do have critical influence on DSC performance. However, the precise control of the binding structure of dyes onto nanocrystalline TiO$_2$ surface to optimize the device efficiency remains a daunting task. In 2013, Zhang et al. successfully manipulated the adsorption structure of Eosin Y dyes on TiO$_2$ substrate by changing the PH value of the organic electrolyte.[28] Figure 2b shows Raman spectra of EY/TiO$_2$ measured under different electrolyte conditions. By adding a small fraction of water into electrolyte, the p$K_a$ value of EosinY carboxyl becomes lower than the pH value of P25 TiO$_2$ system, the hydrogen atom of carboxyl group dissociates and transfers to the solution or to TiO$_2$ surface, leading to an interface structure transition from hydrogen bonded monodentate to bidentate bridging configuration, and enhancing the energy conversion efficiency of the corresponding fabricated photovoltaic device. This work has established a direct link between microscopic interface adsorption structures and macroscopic photovoltaic performance, and has lightened a new way to optimize DSC efficiency by manipulating interface binding configurations.

## 3. ABSORPTION PROPERTY

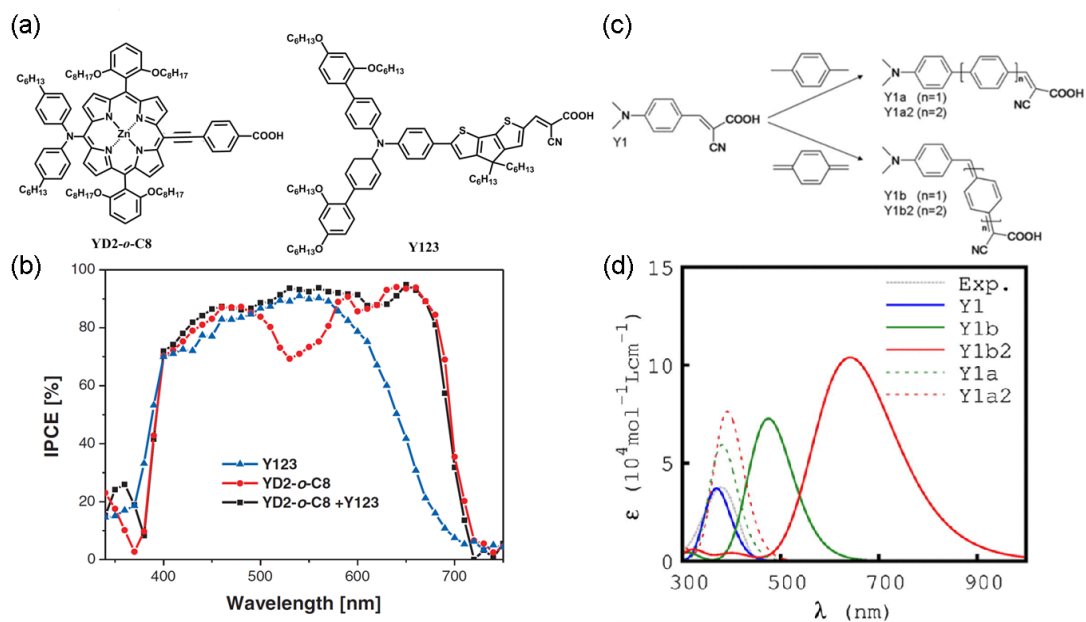

Figure 3. (a) Chemical structures of porphyrin dye YD2-o-C8 and all-organic dye Y123. (b) Spectral response of the IPCE for YD2-o-C8 (red dots), Y123 (blue triangles), and YD2-o-C8/Y123 cosensitized nanocrystalline TiO2 films (black squares).[30] (c) Chemical structures of dyes Y1 and derivatives. (d) Absorption spectra of dye Y1 and derivatives. The grey dashed line is experimental curve for Y1.[32]

Photon absorption by chromophores is the first step in the sequence of processes for energy production in DSC. The optical absorption property of the sensitizer directly determines how much solar energy can be converted into electricity at most. Shockley and Queisser calculated that an ideal absorption threshold energy that absorb photons to produce highest efficiency would be in the range of 1.3-1.4 eV (roughly 940-890 nm),[5] which is readily met by bulk semiconductors that have a direct allowed optical transition at the appropriate band gap energy. However, molecular absorbers always have absorption onsets much higher than the calculated optimum 1.3-1.4 eV. For example, N719, the most commonly used ruthenium bipyridine dye, has an absorption onset at ~1.65 eV (750 nm), significant higher than the ideal threshold. [29]

In recent years, tremendous developments have been made in engineering novel fabrication structures and dyes to improve the corresponding light harvesting property. One way to broaden absorption spectrum is to use co-sensitization of two dyes with complementary absorbance. [3-4,30] For instance, the high DSC power conversion efficiency of over 12% is achieved by using modified Zn-porphyrin dye YD2-o-C8 coadsorbed with metal-free dye Y123 (see Fig. 3a and 3b) by Grätzel et al.. [30] As evidenced by the Incident Photon-to-electron Conversion Efficiency (IPCE) spectra shown in figure 3b, Zn-porphyrin dye YD2-o-C8 lacks absorption in the range 480-630 nm. This dip in the green spectral region can be compensated by dye Y123, which possesses strong absorption capacity around 532 nm. In 2013, Wang et al. achieved a comparable over 11% efficiency using device made via co-grafting of metal free all-organic dye C258 or C259 with dye C239, which have advantages of non-toxicity, easy synthesis, low cost and high extinction coefficients comparing with the metal-based dyes. [4]

Although co-sensitization yields high efficiency in DSC, the fabrication and optimization of these devices can be laborious and technically challenging. The development of a single sensitizer with a panchromatic light harvesting character remains a main objective in the realization of maximum PCEs with standard device fabrication protocols. Most recently, a new dye SM315, with incorporation of the proquinoidal benzothiadiazole (BTD) unit into prototypical structure of D-π-A porphyrins, is reported to achieve unprecedent energy efficiency of 13% at full sun illumination without the requirement of a co-sensitizer. The utility of electron-deficient BTD-functionalized anchor significantly broadens the Soret and Q-band absorbance of porphyrins, yields impressively high light harvesting across the whole visible wavelength range, resulting in an improved $J_{SC}$. [3]

Additionally, another way to improve the sensitizer's light harvesting property is to adjust dye's absorption in the near-IR range of solar irradiance, for example, by replacing dye's donor with stronger electron attributors such as ullazine group [22], or by employing electron rich π-linkers [30-31]. Jiao et al. designed a series of donor-π-acceptor dyes (Y1 and derivatives, see Fig. 3c) with paraquinoid rings as π-conjugation moiety as sensitizers in DSC. [32] The introduced paraquinoid rings drastically shift optical response from violet-blue to near-infrared and significantly enhance photoabsorption of the chromophore, comparing with the small changes in absorption spectrum of donor-π-acceptor dyes with phenyl group as π-bridging unit (see Fig. 3d). In addition, real time excited state electron dynamics simulations based on time-dependent density functional theory indicate that these paraquinoid conjugation dyes maintain high thermal stability when adsorbed on TiO$_2$ surface and ultrafast electron-hole separation at ambient temperature. This simple, while effective, infrared dye Y1b2 is predicted to reach high energy conversion efficiency close to 20% in ideal theoretical conditions[32].

4. **ELECTION INJECTION**

After photo-excitation, the electron-hole separation, integral to the functioning of the cell, occurs by electron transfer from the photo-excited chromophore into the conduction band of the nanocrystalline semiconductor, in a time ranging from subpicoseconds [33-35] to tens of picoseconds

[36] and even nanoseconds. Efficient electron injection is fundamental for DSC operation, directly determines the short circuit current of the photovoltaic device. Electrons in excited states after photo-excitation, if not rapidly injected, easily lose the absorbed photon energy as heat through electron-phonon scattering and subsequent phonon dissipation thus generating (thermal) loss in efficiency. Therefore, it is of crucial importance to fully understand the interface electron transfer dynamics both experimentally and theoretically for further development of the nanoparticle-based device.

**4.1 Experimental techniques measuring interface electronic dynamics**

Ultrafast laser spectroscopy, which is by measuring the excited state dynamics of the sensitizer through transient absorption or fluorescence decay, is most popularly used in measuring electronic dynamics between semiconductor nanoparticles and dye sensitizers. Transient absorption spectroscopy, also known as flash spectroscopy, uses an excitation (or pump) pulse (promoting a fraction of the molecules to their electronically excited state) and a weak probe pulse with low intensity to avoid multiphoton/multistep processes and a delay τ with respect to the pump pulse to record information on the interfacial dynamic processes by calculating difference absorption spectrum ($\Delta A$) between the absorption spectrum of the excited sample and the absorption spectrum of the sample in the ground state.

As most transient absorption studies in the visible and near-IR region are hindered by spectral overlap of absorption in various electronic states, such as the excited states, cationic state, and ground state, as well as stimulated emission, there have been many conflicting reports of electron transfer (ET) rates. Femtosecond mid-IR spectroscopy [37] can directly study the electronic dynamics at the adsorbate/semiconductor interface systematically by measuring IR absorptionconsisting of free carrier absorption, intraband transitions between different valleys (or subbands) within the conduction or the valence bands, and trap states absorption. Since the IR absorption of electrons are direct evidence for the arrival of electrons inside semiconductors, they provide an unambiguous spectroscopic probe for studying interfacial electron transfer between the semiconductor and adsorbates.

The time-correlated single photon counting (TC-SPC) technique [38] is also an ideal method which allows multi-wavelength imaging in conjunction with a laser scanning microscope and a pulsed excitation source to investigate excited state lifetime of electrons at interface. The TC-SPC technique is based on a four-dimensional histogramming process that records the photon density over the time of the fluorescence decay, the x-y coordinates of the scanning area, and the wavelength, which has advantages of ultra-high time resolution (25 ps full-width at half-maximum), ultra-high sensitivity (down to the single photon level) and perfect signal-to-noise ratio. It can accurately describe the electron transfer process between the sensitizers and the $TiO_2$ substrate.

**4.2 Theoretical approaches describing interface electronic dynamics**

Empirical theoretical approaches, which are mainly based on optimized structural features, ground-state molecular dynamics simulations, and/or with empirical kinetic parameters (such as assuming an exponential decay of ET rate as a function of dye length [39] and constant electron-phonon coupling strength[40]) , have been commonly employed to deal with the critical electron transfer process at the dye/$TiO_2$ interface. For instance, Persson et al. have studied the influence of anchor-cum-spacer groups on electron transfer time by approximating the effective electronic coupling strength with the calculated band width for heterogeneous electron transfer interactions based on ground-state DFT calculations.[41] Abuabara et al. successfully investigated the influence of temperature changes on electron injection at dye/TiO2 interface using ground-state molecular dynamics and studied the electron transfer process using an extended Hükel Hamiltonian.[42] Prezhdo et al. reproduced injection dynamics of model chromophores with atomistic details using ground state molecular dynamic simulation and time domain non-adiabatic trajectory surface hopping based on ground-state trajectories.[43] Li et al. studied electron transfer from perylene derivatives into the anatase $TiO_2$ (101) surface using density functional theory (DFT) and a Fock matrix partitioning method.[44] Jones et al. could rapidly predict the injection rate in DSC by partitioning the system into molecular and semiconductor subsystems and computing the retarded Green's function.[45]

However, there are some problems associated with these empirical models:

(i) The excited state potential energy surfaces (PES), which are different from ground state PES, are missing in these simulations, thus the electronic properties in excited states cannot be addressed adequately.

(ii) The electronic couplings at the interface, which is subject to molecular details of the dyes and their dynamic binding configurations on $TiO_2$, cannot be described precisely, thus the time scales obtained therein are questionable.

Real-time time-dependent density functional theory (TDDFT),[46] which evolves quantum mechanically the wavefunctions of excited electron-hole pair at the dye/$TiO_2$ interface based on excited state Hamiltonian, has been used to describe the interfacial electronic dynamics and demonstrated to be especially adequate to treat the interface electronic dynamics and yield consistency with experiment by Meng et al.[18, 35, 47-51] This TDDFT approach has advantages over the previous methods in several aspects:

(i) Very efficient atomic orbital basis sets are adopted, which is small in size and fast in performance.

(ii) Either periodic system or a finite-sized supercell with large vacuum space can be treated without heavy calculation cost.

(iii) Real time excited state trajectories with many-electron density self-consistently propagating at every electronic and ionic steps and forces calculated from mean-field theory are achieved.

Therefore, both experimental and theoretical methods offer a promising way to investigate the interface electron transfer dynamics in chromophore/semiconductor system, allowing a systematic study of the dependence of ET rates on the specific properties of the adsorbates, semiconductors, and the solvent environments.

### 4.3 Factors affecting electron injection
### 4.3.1 Bridging length

According to Marcus theory [51,52], the electron injection rate is strongly dependent on the electronic coupling strength and driving force between the sensitizer and the semiconductor substrate. Electronic coupling strength between molecular excited states and $TiO_2$ can be modified through bridging length between the adsorbates and the binding group. Lian et al. have explored the influence of bridge length on interfacial ET rates by measuring ultrafast electron injection into $TiO_2$ from a rhenium complex with $n = 0-5$ methylene spacers inserted between the bipyridine rings and the carboxylate anchoring groups using femtosecond infrared spectroscopy [53]. They have found that the injection rate decreased exponentially with increasing number of spacers.

**4.3.2 Anchoring group**

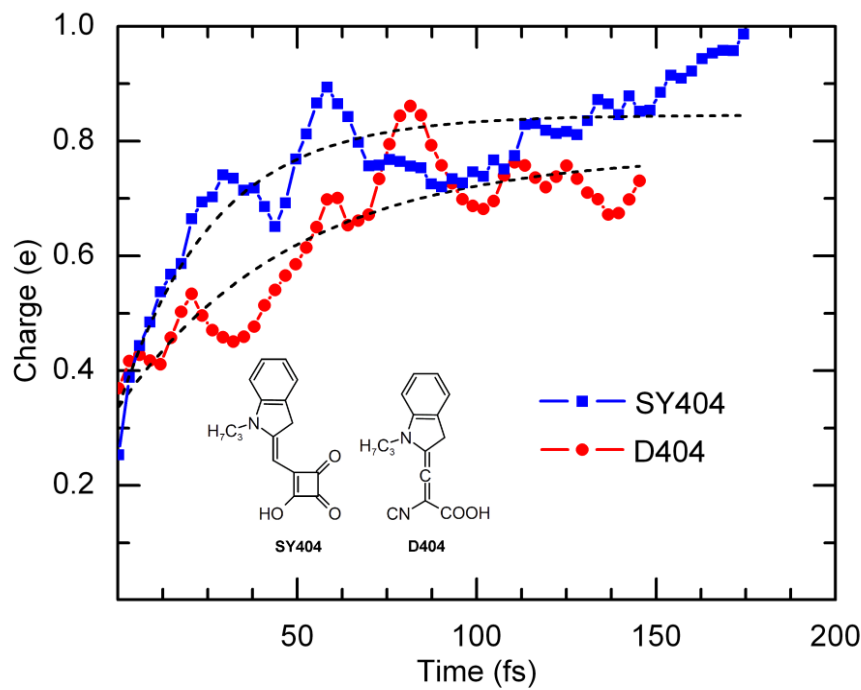

Figure 4. The fraction of photo-excited electrons distributed onto $TiO_2$ substrate as a function of time after photo-excitation of dyes SY404 and D404 with different anchor groups using real-time TDDFT simulation. The insets are chemical structures of dyes SY404 and D404. Dashed lines show exponential fittings of the injection dynamics.

In addition to bridge length, variation of the binding groups is another factor affects the coupling. In 2007, Lian et al. have investigated the effects of anchoring groups on electron injection by comparing ET rate from a ReC1 complex to metal oxides ($TiO_2$, $SnO_2$, ZnO) through carboxylate and phosphonate groups.[54] Faster injection dynamics was observed from phosphonate based chromorphore, which lead to a stronger electronic coupling between bipyridine ligand and metal center than the carboxylate group. Recently, Bartelt et al. synthetized a series of semi-squarylium dyes with a novel acyloin anchor group and investigated the electron injection properties of these dyes using a combination of ultrafast and photoemission spectroscopy.[27] They found this acyloin anchor group show stronger electronic coupling with the substrate and facilitate ultrafast electron injection into $TiO_2$ comparing with a carboxylic acid anchored indoline dye D131. More directly, we have calculated the injection rates of two dyes D404 (with cyanoacrylic acid anchor moiety) and SY404 (with acyloin anchor group) sharing the same donor using real-time excited states TDDFT simulation. From figure 4, at time $t = 0$, electrons have a dominant distribution on the sensitizers, then the energy difference between the molecular LUMO level and the $TiO_2$ conduction band minimum drives electrons efficiently inject into the substrate with a lifetime of 33 fs for dye SY404 and 60 fs for dye D404. Apparently, SY404 dye with acyloin anchor possesses a faster electron injection dynamics due to the stronger electronic coupling strength with the TiO$_2$ substrate.

**4.3.3 Adsorption configuration**

Moreover, even for the same sensitizer, different adsorption configurations would result in different coupling strength, thus leading to different interfacial electron injection. In ref. [18], Jiao et al. have studied how the adsorption structure impacts on the ET lifetime at dye/TiO$_2$ interface using TDDFT electron-ion dynamics simulation. Figure 4a shows the fraction of excited state photoelectron $\chi$ distributed onto TiO$_2$ substrate as a function of time after photo-excitation of systems with binding structures of Ic, IIb, IIc (corresponding to the configurations shown in Fig. 2a). Both the three cases exhibit an ultrafast electron injection process. At time $t = 0$, photo-excited electrons are mainly distributed in the excited states of chromophore and start to inject into the TiO$_2$ conduction band at approximately 16 fs in a exponential way, ultimately finishes at about 100 fs. Exponential fitting shows an injection lifetime of 64 fs for Ic, which slightly slower than the injection time of IIb and IIc (59 fs). Although Ic shows slower injection dynamics, it has a larger quantum yield (equilibrium photoelectron fraction) after injection, 70% comparing to 37% for IIb and IIc, which results from the stronger electronic coupling at the interface in IIb and IIc configurations, leading to substantial state mixing between the dye LUMO and TiO$_2$ conduction band.

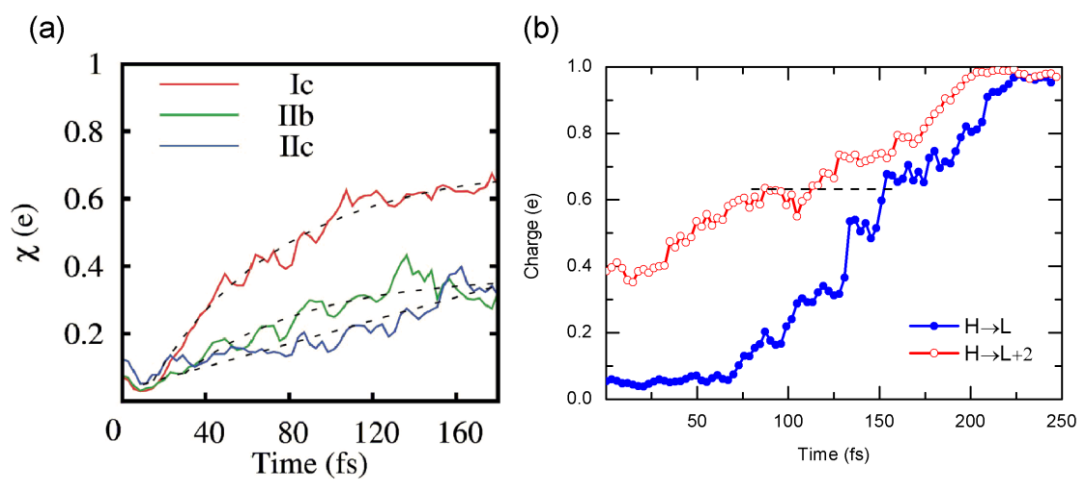

Figure 5. (a) Electron injection dynamics for adsorption configurations Ic, IIb, and IIc from coupled electron–ion MD simulation based on real-time TDDFT. Dotted lines are fitted by an exponential function. (b) The fraction of photo-excited electrons distributed onto TiO$_2$ substrate as a function of time after photo-excitation of dye N1 with different driving forces using real-time TDDFT simulation.

**4.3.4 Driving force**

Besides electronic coupling strength, driving force, which is defined as the potential difference between the molecular excited states and the TiO$_2$ conduction band minimum, is another critical factor influence the interfacial electron injection kinetics. Recently we designed a system using model dye N1 as sensitizer and TiO$_2$ anatase (101) surface as substrate. They investigated the photoelectron injection behaviors from different molecular excited states of the system using real-time excited state TDDFT simulation. At time t = 0, one electron is promoted from the HOMO to higher excited states (LUMO and LUMO+2) of the organic dyes, representing a first excited state of a pair of electron and hole is generated upon photo absorption. As shown in figure 4b, excited electrons are completely injected into the CB of the $TiO_2$ substrate within a time scale of 87 fs from LUMO+2 excited state and 160 fs from LUMO excited state of dye N1, while holes keep stable and confined within the dye molecules. Here the lifetime of the injection process is estimated by the time when 63.2% electrons are transferred from the sensitizer into the $TiO_2$ electrode. Apparently, electrons undergo a faster injection dynamics with larger driving force. In fact, according to Marcus theory, the electron transfer rates are not directly dependent on driving force but rather on the activation energy, which is related to the sum of driving force and reorganization energy. Here, as reorganization energies are same for injections from different excited states of same dye, driving force directly determines the ET rates at N1/$TiO_2$ interface.

## 5. THERMAL RELAXATION

Although larger driving force contributes to ultrafast electron injection, resulting in larger incident photon-to-electron conversion efficiency (IPCE), it brings large energy loss because photon energies in excess of the threshold energy gap are usually dissipated as heat and cannot be converted into electricity. In DSC, an electron is excited from ground state to higher excited state by photon absorption, then injects into the conduction band of the semiconductor substrate, leaving an electron deficiency (hole) in the sensitizer. Those hot electrons generated with photon energy in excess of the HOMO-LUMO band gap quickly cool (within ~1 ps) to the band edges through sequential emission of phonons after injection. Ross et al. have shown that a single-threshold quantum-utilizing device in which the excited carriers thermally equilibrate among themselves, but not with the environment, can convert solar energy with an efficiency as high as 66%. [55]

### 5.1 Hot electron injection

One way to utilize the hot carrier energy is to quickly transfer the hot electrons to the conducting contact before they cool. Hot carrier cooling rates are dependent upon their effective masses and the density of the photo-generated hot carriers (i.e. the absorbed light intensity). Quantization effects in the space charge layer will dramatically slow down the thermal relaxation and enhance the hot electron transfer out of the semiconductor. When the carriers in the semiconductor (i.e. in semiconductor quantum wells, quantum wires, quantum dots, superlattices, and nanostructures) are confined by potential barriers to regions of space that are smaller than or comparable to their deBroglie wavelength or to the Bohr radius of excitons in the semiconductor bulk, the relaxation dynamics can be markedly altered; specifically, the hot-carrier cooling rates may be remarkably reduced. William et al. have successfully slowed down the hot electron relaxation using colloidal PbSe quantum dots as sensitizers and 1,2-ethanedithiol (EDT) passivated nanocrystalline rutile $TiO_2$ substrate, and observed the hot electron transfer from the higher excited states of PbSe quantum dots to the $TiO_2$ substrate within 50 fs using optical second harmonic generation (SHG).[56] Figure 5a shows the temperature-dependent decay of the pump-induced SHG signal. A substantial rise in SHG signal, which is consistent with the hot electron transfer from PbSe to $TiO_2$, is observed after photo-excitation on a time scale shorter than the laser pulse (50 fs). Then a decrease in SHG signal causing by hot electron cooling is observed. The relaxation rate increase exponentially with temperature, resulting in drastic decrease of the SHG signal.

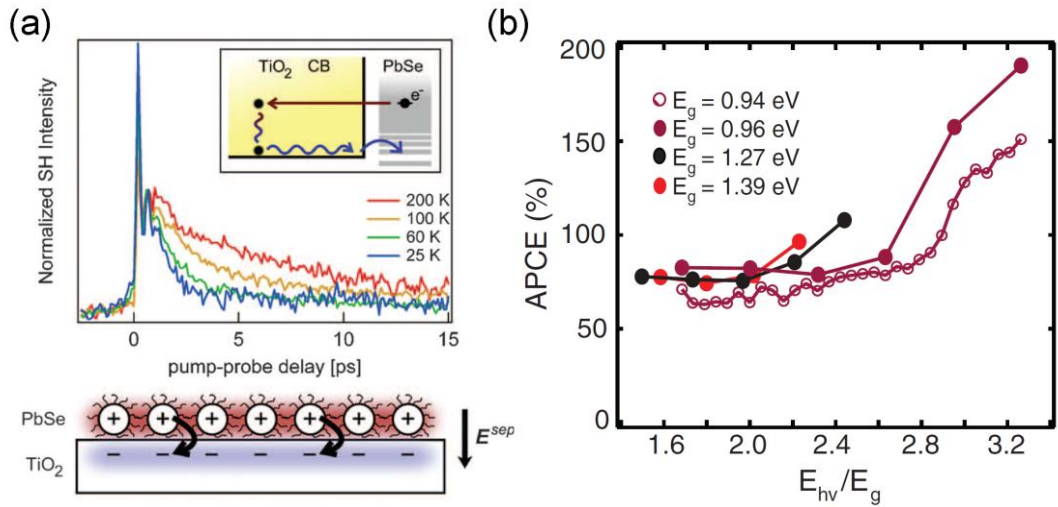

Figure 6. (a) Temperature dependent decay of the pump induced SHG signal enhancement; the absolute intensity has been normalized for pump-induced change to better illustrate the temperature-dependent recovery rate. The bottom figure is the schematic representation of the interfacial electric field generated by separation of electrons and holes across the PbSe-TiO2 interface. [56] (b) APCE versus the incident photon energy divided by the quantum dot band gap energy (indicating the multiples of the band gap). [65]

**5.2 Multiple-exciton generation**

Another effective way to utilize the excessed photon energy is multiple-exciton generation (MEG), which is the creation of two or more electron-hole pairs from one high energy photon by impact ionization. MEG happens on the condition that the rate of impact ionization is greater than the rate of electronic relaxation. MEG has been recognized for over 50 years in bulk semiconductor and been observed in the photocurrent of bulk p-n junctions in Si, Ge, PbS, PbSe, PbTe and InSb by impact ionization. [57-63] However, the threshold photon energy for MEG (where the impact ionization rate is competitive with phonon scattering rates) in bulk semiconductor is many multiples of the band gap, resulting in inefficient photovoltaic output. Fortunately, the impact ionization process is found more efficient in semiconductor nanocrystals or quantum dots because the inverse Auger process of exciton multiplication is greatly enhanced due to quantum confinement effects. Matthew et al. firstly observed the MEG yields in colloidal Si nanocrystals using ultrafast transient absorption spectroscopy in 2007. [64] They found the threshold photon energy for MEG in 9.5 nm diameter Si nanocrystals (effective band gap Eg = 1.20 eV) to be 2.4 ± 0.1Eg and find an exciton-production quantum yield of 2.6 ±0.2 excitons per absorbed photon at 3.4Eg. Parkinson et al. used a photoelectrochemical system composed of PbS quantum dots chemically bound to $TiO_2$ single crystals and demonstrated the multiple-exciton collection (MEC) in experiment for the first time. [65] Figure 5b shows the calculated absorbed photon-to-current efficiency (APCE) values as a function of the ratio of the excitation energy to the band gap of PbS quantum dots. Without MEC process, the APCE values remained constant at ~70% for all the sizes of quantum dots. The APCE values of quantum dot with band gap $E_g$ = 0.94 eV, increase rapidly and exceed 100% at illumination energies larger than 2.7 times of the nanocrystal band gap, indicating the multiple-exciton generation and collection processes. The strong electronic coupling and favorable energy alignment between PbS quantum dots and bulk $TiO_2$ promote the generation and quick collection of multiple excitons from higher excited states.

## 6. ELECTRON-HOLE RECOMBINATION

In addition to thermal relaxation, electron-hole recombination is another main loss of absorbed solar energy. Electrons injected into the TiO$_2$ conduction band, easily lose energy by recombining with holes in electrolyte and the oxidized sensitizers if not rapidly transport to the conducting contact, hence limiting the attainable energy conversion efficiency. Recombination with holes in the oxidized dyes and holes in the electrolyte acceptor species are intertwined together and difficult to separate in experiment. Generally, there are two ways to decrease the electron-hole recombination process in DSC: retarding the recombination with holes in electrolyte and blocking the recombination with holes in sensitizers.

In experiment, people could suppress the electron-hole recombination in electrolyte by changing the size of electrolyte ions or adding additives. For instance, by replacing the traditional I$^-$/I$_3^-$ redox couple with cobalt-complex electrolyte, electron recombination with holes in solution is dramatically reduced, increasing the electron collection efficiency of the device.[3-4, 30] As the size of the cobalt complex is larger than the I$^-$/I$_3^-$ redox couple, it is difficult for cobalt complexes to make contact with the nanocrystal TiO$_2$ surface directly, hence retarding the charge recombination in electrolyte. Adsorption of Li$^+$ ions from the electrolyte on the semiconductor TiO$_2$ surface can also slow down both electron transport and charge recombination remarkably.[66] Dai et al. have found that by introducing a special additive, tributyl phosphate (TBpp), to modify the dyed-TiO2/electrolyte interface, the electron recombination at the dyed-TiO$_2$/electrolyte interface was restrained and the photovoltaic performance was enhanced by 40%.[67] The TBpp parent molecule split into several smaller fragments and form four anchoring modes on the TiO$_2$ surface. The molecular cleavage of TBpp and adsorption of N719 assist each other on the sensitized TiO$_2$ surface, transforming the unstable N719 configuration into stable N719 configuration, thus reducing N719 aggregation at the dye/TiO$_2$ interface. Furthermore, these new fragments are multiply adsorbed on the non-sensitized TiO$_2$ surface to form an insulating barrier layer. Therefore, the interface electron-hole recombination is retarded.

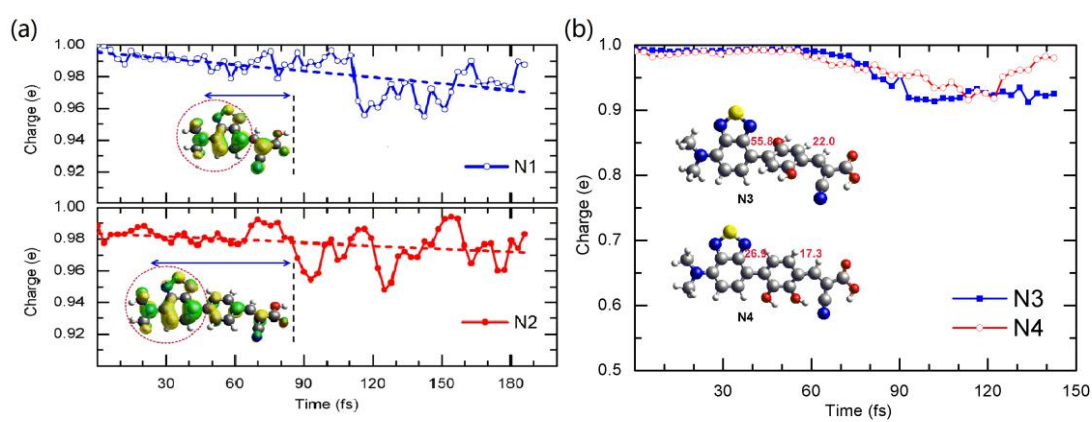

Figure 7. (a) Fraction of electrons transferred from the TiO$_2$ semiconductor substrate to the organic dyes N1 and N2 after excitation at the organic dye-TiO$_2$ interface. Dashed lines are results fitted by a linear decaying dynamics. The two insets show back electron transfer distances from the semiconductor TiO$_2$ to the sensitizers. (b) Fraction of electrons transferred from the TiO$_2$ semiconductor substrate to the organic dyes N3 and N4 after excitation at the organic dye-TiO$_2$

interface. Insets are chemical structures of N3 and N4. Dyes N3 and N4 are isomers with different dihedral angles between the donor moiety and the bridging unit (shown by numbers therein).

On the other hand, making slight structural modifications to dye molecules can also slow down the charge recombination in oxidized dyes. Haid et al. have decreased electron-hole recombination rates by about 5 times through inserting a phenyl ring between the benzothiadiazole (BTDA) bridging unit and the cyanoacrylic acid acceptor.[36] Ma et al. have found that the five times changing in recombination rate mainly come from the longer back electron transfer distance of the inserted dye using quantum chemical simulations.[49] As shown in figure 7, they used dye N1 and N2 as simplified model of dye 1 and dye 2 in ref. [36] to investigate how the small structural modification have significant influence on charge recombination. By the insertion of an additional phenyl ring close to the anchoring group, the electron-hole recombination rate is slowed down by about four times (23 ps vs. 6 ps). Charge transfer distance dependence is found to be the main factor for this significant difference in the recombination lifetime by theoretical analysis.

Besides longer recombination distance, many people think that structural twisting of donor-π-acceptor dye can break down the π-conjugation between the donor and acceptor and thus block electron back transfer to the sensitizer from the charge separated state. We have investigated the influence of structural twisting on dye/$TiO_2$ interface electron-hole recombination by calculating the recombination dynamics of two isomers dyes N3 and N4 adsorbed on $TiO_2$ anatase (101) surface based on real time excited state TDDFT. Figure 7b shows the evolution of electrons transferred back from the $TiO_2$ conduction band to the sensitizers. Two insets are the chemical structures of dyes N3 and N4. N3 and N4 share same compositions but have ~ 30 ° difference in dihedral angles between the donor moiety and bridging unit. Obviously, dyes N3 and N4 exhibit similar recombination dynamics and the planar dye N4 even shows a slower recombination dynamics. Therefore, structural torsion of organic dyes hardly affects the recombination process at dye/$TiO_2$ interface.

7. **CONCLUSION AND OUTLOOK**

In this review, we briefly introduce the composition, working principles, and recent progresses of dye-sensitized solar cells, with a special focus on atomistic level information obtained from recent extensive investigations. From the five critical factors affecting the solar cell efficiency, including dye/$TiO_2$ interface structures, dye absorption properties, electron injection, thermal relaxation and electron-hole recombination, we introduce the basic concept of these factors and the roles they play in DSC, the advanced theoretical and experimental methodologies, the influence of these individual processes on overall DSC efficiency, past achievements and future opportunities for further improvements.

However, improving the DSC efficiency is a grand challenge which requires both breakthroughs in fundamental concepts and finest systematic engineering. There exists no universal method for all cases chasing for a better photovoltaic performance. For instance, near IR dyes have better photon absorption, but its $V_{OC}$ is relatively low. Adding additives of TBpp can on one hand slow down the electron-hole recombination at $TiO_2$/electrolyte interface, but on the other hand, it also slows down electron transport rates in $TiO_2$ nanoparticles. MEG and MEC processes have been only achieved in quantum dots solar cells, but not been available in organic DSC.

Clearly, the efficiency of DSCs is still far from the Shockley-Queisser limit for a single absorber, great efforts should be made in more focused researches and systematic methods. Several promising approaches can be taken to boost further PCE efficiency, for example, design of novel dyes with panchromatic absorption properties and large molar coefficient to significantly enhance light harvesting efficiency; or design of dye structures to hinder the approaching of redox species to the $TiO_2$ surface, thereby lowering the rate of back electron transfer. In addition, fundamental research with precise characterization on the specific interface structures and dynamic processes is of crucial importance for future device optimization. Scanning tunneling microscopy and spectroscopy, for instance, are demonstrated excellent in describing structural and electronic properties of various dye molecule geometries at interface, with a single-molecule resolution.[68] Non-contact atomic force microscopy with a functionalized tip might be effective in directly imaging dye adsorption on non-conductive substrates.[69] More importantly, with extraordinary field enhancements under a sharp metal tip, detection of ultrafast electron dynamics for individual dye molecules or dye configurations might be possible. For realistic large-scale implementation, cost and stability are another two major preoccupations in DSC research.

## ACKNOWLEDGEMENT

We acknowledge financial supports from the NSFC (grants11222431 and 11074287), the MOST (2012CB921403), and the hundred-talent program of CAS.